\documentclass[aps,pra,reprint]{revtex4-1}
\usepackage{hyperref}
\usepackage{amsmath}
\usepackage{amssymb}
\usepackage[pdftex]{graphicx}
\usepackage{microtype}
\usepackage{verbatim}
\usepackage{todonotes}
\presetkeys{todonotes}{inline,backgroundcolor=yellow}{}
\usepackage{subcaption}
\makeatletter
\DeclareRobustCommand{\element}[1]{\@element#1\@nil}
\def\@element#1#2\@nil{%
  #1%
  \if\relax#2\relax\else\MakeLowercase{#2}\fi}
\makeatother

\newcommand{\mrm}{\mathrm}

\begin{document}
%\linenumbers
\title{Cold, dense atomic ion clouds produced by cryogenic buffer gas cooling}
\author{Nishant Bhatt}
\author{Kosuke Kato}
\author{Amar C. Vutha}
\email{vutha@physics.utoronto.ca}
\affiliation{Department of Physics, University of Toronto, Toronto, Canada M5S 1A7}

\begin{abstract}
We produce cold and dense clouds of atomic ions (Ca$^+$, Dy$^+$) by laser ablation of metal targets and cryogenic buffer gas cooling of the resulting plasma. We measure the temperature and density of the ion clouds using laser absorption spectroscopy. We find that large ion densities ($\gtrsim 10^9$ cm$^{-3}$) can be obtained at temperatures as low as 6 K. Our method opens up new ways to study cold neutral plasmas, and to perform survey spectroscopy of ions that cannot be laser-cooled easily.
\end{abstract}
\maketitle

\section{Introduction}
Precision spectroscopy of ions has led to numerous advances and applications. Optical clocks based on trapped ions are the most accurate clocks to date \cite{Huntemann2016,Dube2016,Chen2017,Brewer2019}. Trapped ions are one of the leading approaches to the development of large-scale quantum processors \cite{Haffner2008,Monroe2013,Debnath2016,Bruzewicz2019}. Precision measurements using atomic \cite{Quint2011,NunezPortela2013,Godun2014} and molecular ions \cite{Bressel2012,Cairncross2017} have advanced our understanding of standard model and beyond-standard-model physics. 

However, many potential applications involving atomic and molecular ions are stymied by the limited availability of spectroscopic information. For example, astronomical observation of the signature spectral lines of ions found in space can be compared to laboratory-based measurements to test fundamental physics \cite{Murphy2013,Songaila2014,Campbell2016,Brunken2014} -- however, accurate rest frequencies of complex ions are often unavailable if they have not been laser-cooled, or trapped and sympathetically cooled using another laser-coolable ion. Another example is furnished by molecular ions: there are a variety of fundamental physics experiments that take advantage of the vibrational and rotational degrees of freedom of molecular ions \cite{Bressel2012,Kajita2016,Cairncross2017,Carollo2019}. But the extreme dearth of spectroscopic information about the majority of molecular ions makes it a challenge to work with hitherto unstudied species. 

% Significant advances in spectroscopic techniques have resulted from attempts to address this problem. Sympathetic cooling and quantum logic spectroscopy \cite{Chou,Schmidt} have been used to make extremely precise frequency measurements on atomic and molecular ions that can be co-trapped with laser-coolable ions. However these methods, which typically involve just one spectroscopy ion (and therefore have long cycle times), are best suited for high precision spectroscopy of lines whose positions are known to sufficient accuracy, rather than as survey techniques to map out large frequency ranges. For wideband frequency surveys, an instrument using cavity-enhanced velocity modulation spectroscopy along with a frequency comb has been used for spectroscopy of HfF$^+$ and ThF$^+$ ions \cite{Gresh2016}. 

The spectroscopy of such unexplored atomic and molecular ions would be greatly aided by the availability of a simple and general method for survey spectroscopy of cold ions. Here we adapt a technique that has been widely used for neutral atom and molecule spectroscopy, cryogenic buffer gas cooling \cite{Hutzler2012,Hemmerling2014}, to address this need. We report the first observation of cold and dense samples of atomic ions produced using cryogenic buffer gas cooling of laser-ablation plasmas. We perform absorption spectroscopy on Ca$^+$ ions cooled using helium buffer gas. Additionally, as a demonstration of the applicability of this technique to the spectroscopy of a complex ion, we present measurements made on cold Dy$^+$ ion clouds.

\begin{figure}[h!]
\includegraphics[width=\columnwidth]{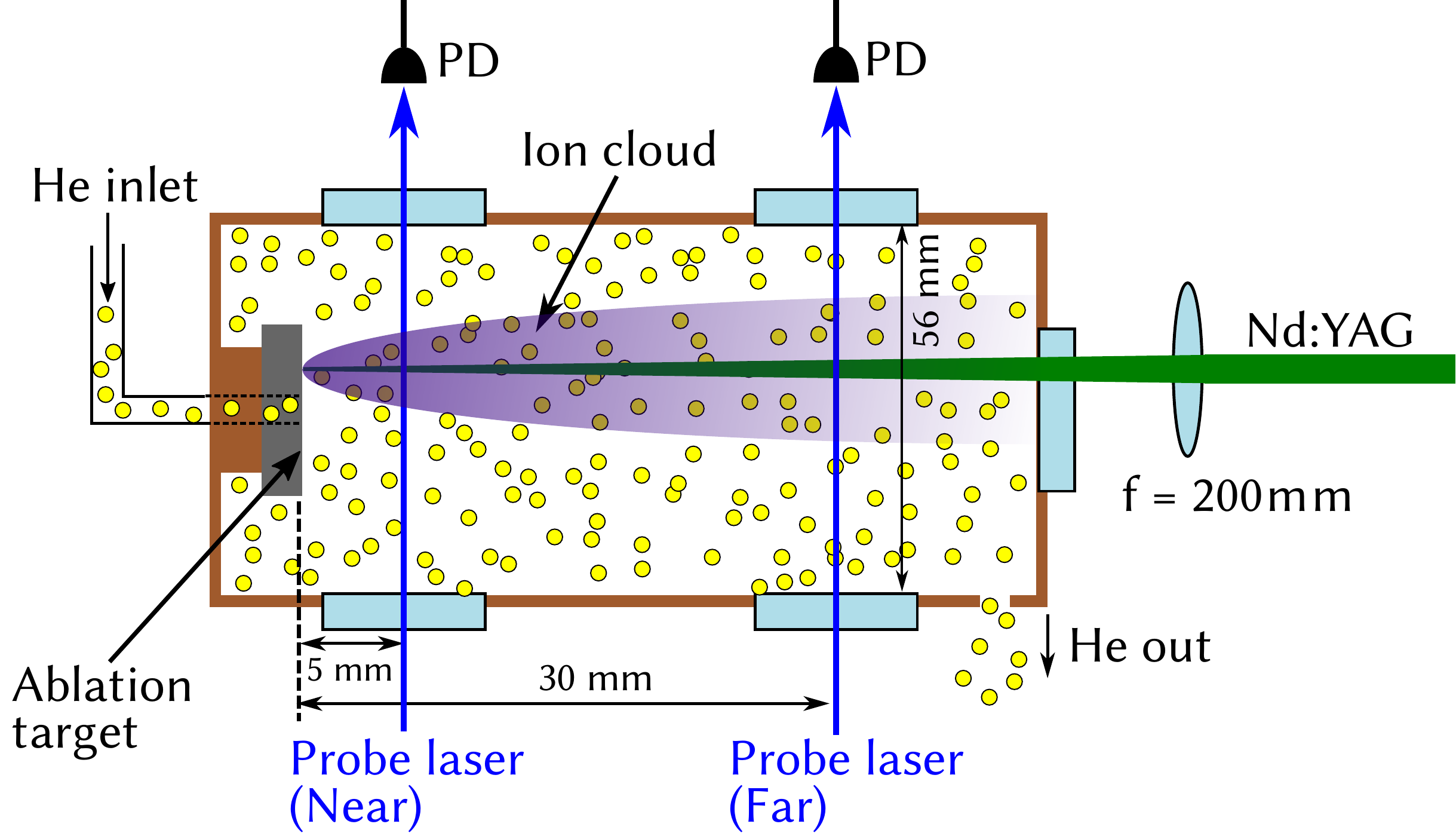}
\caption{Schematic of the cell used for the laser absorption measurements (not to scale). The cell is thermally connected to a liquid helium cryostat, and was maintained at 6 K for all the measurements reported here. PD are photodiodes.}
\label{fig:apparatus}
\end{figure}

\section{Apparatus}
We produced and studied Ca$^+$ and Dy$^+$ ion clouds in a buffer gas cell held within a liquid-helium-cooled cryostat. Calcium and dysprosium metal targets were affixed in an aluminum cell attached to the cryostat. The metal targets were ablated using a pulsed Nd:YAG laser ($\sim7$ ns pulse width) focused on to the targets using a 200 mm focal length lens, with typical pulse energies between 5-10 mJ. (Higher temperatures for the ion clouds were observed with higher pulse energies.) The ions produced by laser ablation reached equilibrium through collisions with helium buffer gas. The buffer gas was pre-cooled to 4 K and introduced into the cell at a flow rate of 1 standard cubic centimeter per minute ($4.5 \times 10^{17}$ atoms/s). Helium gas was allowed to leak out of the cell through a 1 mm diameter aperture, resulting in a steady state buffer gas density $n_\mrm{bg} \approx 2 \times 10^{16}$ cm$^{-3}$ within the cell. 

A schematic of the cell and the laser geometry used for the measurements is shown in Fig.\ \ref{fig:apparatus}. Ca$^+$ ions were probed using the $4s ~^2S_{1/2} \to 4p ~^2P_{1/2}$ transition, and Dy$^+$ ions using the $4f^{10} \, 6s_{1/2} (J=17/2) \to 4f^{10} \, 6p ~ (J=15/2)$ transition (both near 397 nm). The frequency of the 397 nm probe laser was locked (typically with $\lesssim$1 MHz instability) to a cesium-stabilized reference laser using a transfer interferometer \cite{Jackson2018}, and scanned under software control. Photodiode signals were processed to extract the optical depth at both the ``near'' and ``far'' measurement ports shown in Fig.\ \ref{fig:apparatus}. Higher ion densities and temperatures (both typically $\sim$50\% higher) were observed at the near port. All the measurements reported here were made at the far port.

\section{Results}
\begin{figure}[h]
    \centering
    \includegraphics[width=\columnwidth]{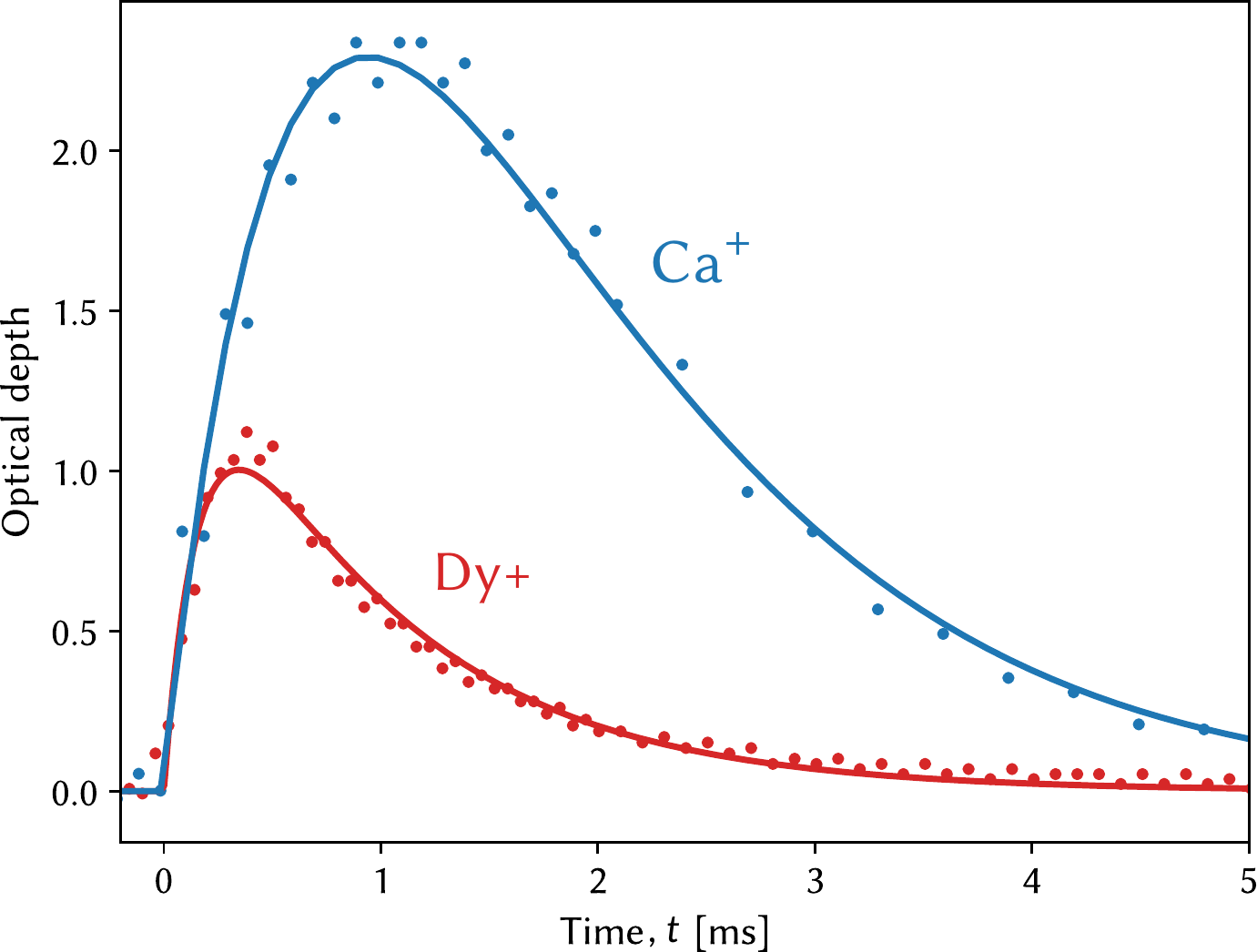}
    \caption{Time evolution of the optical depths of the Ca$^+$ and Dy$^+$ ion clouds. The ablation laser pulse occurs at $t=0$. The solid lines are a fit to the function $F(t)$ described in the text. }
    \label{fig:absorption_time_trace}
\end{figure}

Optical depths were calculated from the absorption time traces measured on the photodiodes, and fit to a simple empirical model function $F(t) = A \left(1 - e^{-t/\tau_1} \right) e^{-t/\tau_2}$, as shown in Fig.\ \ref{fig:absorption_time_trace}. We calculate the resonant absorption cross section $\sigma_0 = 
\frac{\lambda_0^2}{2\pi} \left(\frac{2J_f+1}{2J_i+1}\right) \eta$, where $\lambda_0$ is the resonance wavelength, $J_i,J_f$ are the angular momenta of the initial and final states, and $\eta$ is the ratio of the partial linewidth of the transition to the total linewidth of the excited state \cite{Budker2008}: the cross section for the Ca$^+$ (Dy$^+$) transition is $\sigma_\mrm{Ca^+} = 2.8 \times 10^{-10}$ cm$^2$ ($\sigma_\mrm{Dy^+} = 1.8 \times 10^{-10}$ cm$^2$). Using the relation $n = F/\sigma_0 \ell$, the peak optical depth $F_\mrm{max}$ = 2.2 (1.0) for Ca$^+$ (Dy$^+$) shown in Fig. \ref{fig:absorption_time_trace} corresponds to a peak ion density (averaged along the $\ell$ = 56 mm long absorption column) of $n_\mrm{Ca^+} = 1.4 \times 10^{9}$ cm$^{-3}$ ($n_\mrm{Dy^+} = 10^{9}$ cm$^{-3}$). 

We were able to increase the peak optical depths of the clouds compared to the values shown in Fig.\ \ref{fig:absorption_time_trace} by a factor of $\sim$2 using higher ablation laser pulse energies, at the expense of increased temperature. The fall time, from the fit to the time traces shown in Fig.\ \ref{fig:absorption_time_trace}, is $\tau_2$ = 0.9 ms (0.2 ms) for Ca$^+$ (Dy$^+$) ion clouds. In comparison, the fall time of neutral Ca atom clouds was measured to be significantly longer ($\tau_2\sim$70 ms) with the same buffer gas density and cell geometry, consistent with loss by diffusion to the cell wall and extraction through the cell aperture, as seen in other neutral atom buffer gas cooling experiments cf.\cite{Hutzler2012,Hemmerling2014}.

\begin{figure}[h!]
    \centering
    \includegraphics[width=\columnwidth]{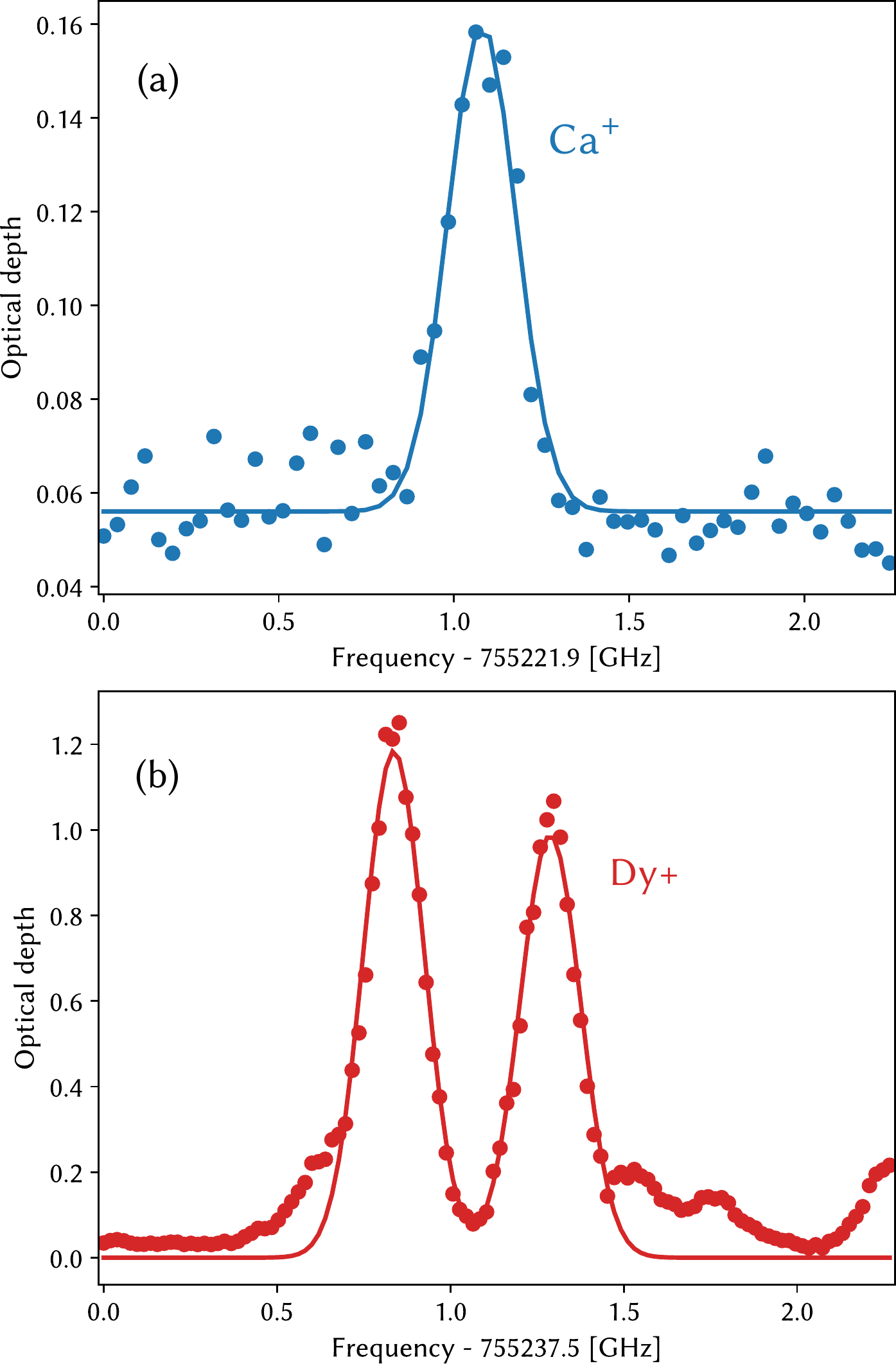}
    \caption{Absorption spectrum of the the ion clouds. The peak optical depths from traces such as in Fig.\ \ref{fig:absorption_time_trace} are measured while the frequency of the probe laser is scanned. The solid lines are fits to a gaussian (sum of two gaussians) for Ca$^+$ (Dy$^+$) corresponding to a Doppler temperature of 6 K.}
    \label{fig:doppler}
\end{figure}
%(The Ca$^+$ spectrum in (a) was acquired with lower density ion clouds to limit saturation of the detectors and to prevent heating of the ions.)

The absorption spectrum of the ion clouds is shown in Fig.\ \ref{fig:doppler}, along with the fits used to extract the temperature of the clouds. All the lines yield identical Doppler temperatures ($T_D$ = 6 K), equal to the cell temperature measured using a calibrated thermometer. The mean time between collisions of the ions with the helium atoms is $\tau_\mrm{c} \sim1 ~\mu$s at the buffer gas density used in these experiments. The long dwell times ($\tau_2 \gg \tau_\mrm{c}$) and low temperatures measured in Figs.\ \ref{fig:absorption_time_trace} and \ref{fig:doppler} indicate that ions from the laser ablation plasma survive a large number of collisions as they thermalize with the helium buffer gas atoms. This is consistent with the expected absence of charge exchange collisions with the buffer gas, given the large ionization energy of helium compared to the electron affinities of Ca$^+$ and Dy$^+$.

The mean ion spacing at the densities measured above is $a = \left({3}/{4\pi n_\mrm{ion}} \right)^{1/3} \approx$ 6 $\mu$m, which translates to a pairwise Coulomb repulsion energy $V_\mrm{C} = \frac{1}{4\pi\epsilon_0} \frac{e^2}{a} \approx k_B \times$ 3 K. This is a significant amount compared to the thermal energy of the ion clouds, indicating that the ions along with an accompanying electron cloud exist as a neutral plasma in the buffer gas cell. Compared to typical ultracold neutral plasmas produced by photoionization of laser-cooled neutral atoms (cf.\ \cite{Lyon2016,Langin2019}), we observe at least an order of magnitude more cold ions at comparable densities, confined within the buffer gas cell for a longer duration. The Coulomb coupling parameter for the Ca$^+$ and Dy$^+$ ion clouds is $\Gamma_\mrm{ion} = {V_\mrm{C}}/{k_B T_D} \approx 0.5$, which puts them on the verge of the strong-coupling regime \cite{Killian2007,Lyon2016}.

We offer two possible hypotheses for the disappearance of ion signals at a faster rate than neutral atoms under identical conditions: (a) diffusion of electrons to the walls of the cell at a faster rate than ions, followed by Coulomb explosion of the positive ion cloud; or (b) recombination of the electrons and ions into neutral atoms. We consider (b) to be less likely, but it cannot be ruled out conclusively at this time. 

The two Dy$^+$ peaks shown in Fig.\ \ref{fig:doppler} are spaced apart by $\Delta \nu$ = 451 $\pm$ 5 MHz, where the uncertainty is the quadrature sum of the fit uncertainty and the probe laser frequency stability. These two peaks were the largest features in this frequency neighborhood, suggesting that they are $I=0$ isotope lines that are undiluted by hyperfine splitting. Further, the ratio of the heights of these peaks is $r$ = 1.20 $\pm$ 0.06, consistent with the natural abundance ratio of $~^{164}$Dy and $~^{162}$Dy ($r=1.11$). (Only one high-resolution spectroscopic study of Dy$^+$ ions has been published \cite{DelPapa2017}, but that work does not provide hyperfine constants or isotope shifts for the $J=17/2 \to J=15/2$ Dy$^+$ line studied here). We chose to study this Dy$^+$ line due to its convenient overlap with an existing 397 nm probe laser in our laboratory -- nevertheless, the ease with which an unexplored line in a complex ion such as Dy$^+$ can be spectroscopically studied using buffer gas cooling is indicative of the wider applicability of this technique.

\section{Summary}
We have shown that high densities of cold atomic ions can be produced by cryogenic buffer gas cooling. We observe that the ions survive multiple collisions with helium buffer gas atoms and thermalize with them, making it possible to study their optical transitions with good spectral resolution. As with neutral atoms and molecules, buffer gas cooling of ions is a convenient and general method that can likely be used to study a vast array of atomic and molecular ions -- any ions that do not undergo charge transfer reactions with the helium buffer gas can be cooled and studied in this way. This method could be useful for survey spectroscopy of atomic and molecular ions that are interesting for astrochemistry and precision measurements. Cooling and confinement of ions without externally applied electromagnetic fields could also be useful for characterizing collisional frequency shifts relevant for trapped-ion clocks \cite{Davis2019,Hankin2019}, and for studying the many-body physics of strongly coupled neutral plasmas \cite{Killian2007,Lyon2016}.

% \section*{Acknowledgments} 
~\\ \emph{Acknowledgments. --} We acknowledge the contributions of Luca Talamo to the design of the cryogenic system. We are grateful to Prof.\ John Doyle for the loan of cryogenic equipment. We acknowledge funding from NSERC, the Canada Foundation for Innovation and the Ontario Research Fund. ACV acknowledges support from the Canada Research Chairs program and the Sloan Research Fellowship.

\bibliography{cold_dense_ions.bib}

\end{document}